\documentclass[10pts,showpacs,preprint,aps]{revtex4}
\usepackage{graphicx}
\usepackage{dcolumn}
\usepackage{bm}
\usepackage{amssymb}
\usepackage{amsmath}
\begin{document}
\setcounter{page}{1}
\vskip 2cm
\title
{The connection between multiple prices of an Option at a given time with single prices defined at different times: The concept of weak-value in quantum finance}
\author
{Ivan Arraut$^{(1)}$}
\author
{Alan\;Au$^{(1)}$}
\author
{Alan\;Ching-biu\;Tse$^{(1, 2)}$}
\author
{Carlos Segovia$^{(3)}$}
\affiliation{$^{(1)}$ Lee Shau Kee School of Business and Administration, The Open University of Hong Kong, 30 Good Shepherd Street, Homantin, Kowloon}
\affiliation{$^{(2)}$ Department of Marketing, The Chinese University of Hong Kong, Cheng Yu Tung Building 12 Chak Cheung Street Shatin, N.T., Hong Kong}
\affiliation{$^{(3)}$IMUNAM, Oaxaca, Mexico, Leon 2, Col. Centro, Oaxaca de Juarez, Oaxaca, 68000, Mexico.}

\begin{abstract}
We introduce a new tool for predicting the evolution of an option for the cases where at some specific
time, there is a high-degree of uncertainty for identifying its price. We work over the special case where we can predict the
evolution of the system by joining a single price for the Option, defined at some specific time with a pair
of prices defined at another instant. This is achieved by describing the evolution of the system through a financial
Hamiltonian. The extension to the case of multiple prices at a given instant is straightforward. We also explain how to apply these results in real situations.
\end{abstract}
\maketitle 

\section{Introduction}    
The weak-value was introduced for first time in \cite{1}, in order to find a symmetrical formulation of Quantum Mechanics. It is in some sense an extension of the concept of the
S-matrix coming from Quantum Field Theory (QFT), since it joins initial and final states
\cite{2}. This concept (weak-value) is used for analyzing systems where we wish to make some
measurements without affecting their evolution \cite{3}. The weak-value of an operator is in
general a complex number, then it is not comparable to an eigenvalue in ordinary Quantum
Mechanics with a Hermitian Hamiltonian. The concept has been used for finding an alternative explanation of the double slit experiment in Quantum Mechanics \cite{4}. In general, when
we use the concept of weak-value, we have to pre-select one state, which can be interpreted
as an initial condition and we have to post-select another state which can be interpreted as
a final condition. Both states are joined through the evolution described by a Hamiltonian.
In this paper we expect to extend these ideas to the world of the financial market in order to create a powerful predicting tool. The use of Quantum Mechanical tools and the application of them inside the financial market is what is commonly known as "Quantum Finance" \cite{5}. This formulation appears naturally for the cases where the equations explaining the evolution of the price of an Option, can be expressed in the form $\hat{H}\psi=E\psi$, which looks like (but it is not) the Schr\"odinger equation. This provides the possibility of solving the system of equations in the same form as in the case of ordinary Quantum Mechanics, simplifying then the way how we understand the evolution of an Option. In ordinary Quantum Mechanics, the Hamiltonian ($\hat{H}$) in standard situations is Hermitian, such that its eigenvalues are real numbers which guarantee an unitary evolution. However, there are many physical situations where non-Hermitian Hamiltonians have been explored in physics. In \cite{Ellis}, it has been demonstrated that the unitarity for these cases can be recovered if the Hamiltonian is invariant under the simultaneous action of Parity and time reversal (PT symmetry) operations. Other interesting applications for non-Hermitian Hamiltonians have been found before \cite{Ellis2}. In Quantum finance, the emerging Hamiltonians are usually non-Hermitian and we should not expect the same behavior of ordinary Quantum Mechanics. Then the probability is not necessarily conserved in time (the evolution can be non-unitary) since we cannot expect the $PT$ symmetry to be obeyed neither. Note that the fact that the probability is not conserved in time, does not mean that it is not well-defined. In fact, the probability is always well-defined at each instant and its total value is equal to one. The non-conservation of probability only means that the normalization factor (which makes the probabilities well-defined) changes in time. This is a consequence of the fact that the Financial market is not an isolated system, but it is rather an open system with permanent input and output of information. It is well-known that Open systems are modeled with non-Hermitian Hamiltonians and this does not violate any Quantum principle \cite{Newagain, Newagain3, NA4, NA5, Newagain2, kato}. Given the non-Hermitian character of the system (without any imposed symmetry), the eigenvalues of the financial Hamiltonians are in
general complex numbers and the evolution of the system is not necessarily unitary. As
a consequence of this, the kets not necessarily evolve as a plane wave when the solutions
evolve freely (zero potential). In fact, in general cases, the kets evolve as a damping (or
unstable) oscillation with finite or (in some cases) infinite wavelength. The known techniques in Quantum Finance, analyze the dynamics involved in the stock market by using kets evolving backward
in time \cite{5}. This means that we have to exchange the roles
of the pre-selected and the post-selected states in the weak-value formulation if we want to describe the financial systems
by using the weak-value approach. Under the weak-value
approach, the situation described by a double slit experiment is related to the values taken
by the price of an Option at a given time. For the financial case, both states (pre/post selected) are
connected through the time evolution described by the non-Hermitian Hamiltonian. In this
paper we evaluate two scenarios: 1). In the first scenario, we pre-select a single price
and describe its evolution toward a post-selected state defined by two prices for the Option at a single time.
2). In the second case, we pre-select the state defined by two-prices at a given initial time
and then connect it with a single final price defined at a final time for the same Option. Here we also explain how these ideal cases can be connected to the reality. The paper is organized
as follows: In Sec. (\ref{Rev}), we make a review of some fundamental concepts in finance. In Sec. (\ref{BSEAMA}) we derive the Black-Scholes equation which will be used as an example for applying the formulation proposed in this paper. We also express it as an eigenvalue problem after doing an appropriate change of variables. In Sec. (\ref{II}), we make a brief review of the standard formulation of the double-slit experiment in Quantum Mechanics. In Sec. (\ref{III}), we make the analysis of the same
experiment from the perspective of the concept of weak-value. In Sec. (\ref{IV}), we extend the
arguments of the weak-value formulation of the double slit experiment in order to analyze
the behavior in time of the prices of an Option. In this case we connect two initial prices with a single final price of the Option. We then develop an example based
on the Black-Scholes Hamiltonian and we then formulate the generic result valid for any Hamiltonian (any equation). We also explain in this section how useful is the concept of Weak-Value for analyzing real situations. In Sec. (\ref{V}) we again
develop the financial weak-value version of the double slit but this time taking two prices for the final value of the Option, and joining them with a single initial price through
the time-evolution defined by the corresponding financial Hamiltonian (inverse double slit). In Sec. (\ref{Chopra2}), we explain in words how the double-slit approximation, together with the important concept of weak-value, can be applied for the cases where the "Alpha predator" stocks, namely, stocks with high relative trading volume through which high-incomes can be obtained, are being explored. Finally in Sec. (\ref{VI}), we conclude.

\section{Review on concepts related to financial markets}   \label{Rev}

Finance is normally an empirical discipline and this includes the predictions in the Stock market, where people normally prefer to evaluate the situations by intuition and/or collecting data, instead of using analytical approaches. This is the case because normally the industry is in hurry for finding solutions to problems for which it would take long time to find analytical solutions. In addition, the amount of data involved, can be so large that we cannot claim the analytical solutions to be exact. Being empirical can perhaps solve problems in the short term, however, it does not solve long term problems. This is where the analytical analysis becomes important. Different methods present different advantages. In Finance, we have to deal with Financial assets (also called Security), which are pieces of paper which the holders keep in order to claim a fraction of real assets, and the money (if there is any) generated by the real asset. The specific form of a financial asset is called {\it Financial Instrument}. The stock is an example of the financial instrument. From the perspective of the parties providing the Financial asset, it is perceived as a Financial Liability. This is the case because the Financial Instrument has to be divided among all the stockholders. There exist fundamentally three primary forms of financial instruments. They are:\\
1). Equity: These are the usual stocks and shares, which a company distribute. Whoever has a stock in his/her hands, is partially owner of the company. There is no guarantee in receiving an income in this case, unless the company is profitable. Then the owner of a stock is exposed to the same risk of the company. \\
2). Fixed income securities: They are also known as bonds, which are issued by corporations and governments. They guarantee two possibilities, namely, a). A single and fixed payment. b). A stream of fixed payments.\\
The bonds are instruments of debt; this means that the the person who buys a bond is giving a loan to the organization which sells it. Then there is an agreement about the period after which there is a repayment of the debt. The time interval after which the debt is payed is called {\it Maturity}.\\
3). Derivative securities: They are financial assets that are derived from other financial assets.\\

More complex instruments, appear from the combination of the previous ones but we will not analyze them in detail in this paper. 

\subsection{What is a financial market}

A market is defined as the media where the transaction between buyer and seller is done. It can be organized primarily as\\\\
1). Direct market: This consists on the direct search of the buyer and the seller.\\
2). Brokered market: Is the market where the brokers operate as intermediate between the the buyers and the sellers.\\
3). The auction market: This is one where the the buyers and sellers have a simultaneous interaction inside a centralized market \cite{New1}.\\

A market is in equilibrium if there are no more transactions on it since the securities will have reached their "Fair price". This condition is known as {\it efficient market} and it is tied with the amount of information available by the buyers and sellers. When both parties have the same amount of information and wealth, then the equilibrium condition should be reached since they will not want to trade anymore. The equilibrium condition is broken when new (perhaps incomplete and random) information enters the system. In such a case, the system moves toward a new equilibrium when again all the participants in the market have the same amount of information. Even in equilibrium, there might be random changes for the prices for the securities and it is for this reason that these changes are normally represented by random variables \cite{5, 23}. Then given a security $S(t)$, its change is a random variable represented by $\frac{dS(t)}{dt}$.　If $t$ represents the ordinary time, then the security variable itself is random. The randomness of the security $S(t)$ is represented by the volatility parameter $\sigma$. Then for example a large random fluctuation, is represented by a large value of the random parameter $\sigma$. In the limit $\sigma\to0$, then the security has no uncertainty in its future evolution. \\
Financial markets are divided in\\
{\it 1). Capital markets:} These are structured to trade with the instruments of equity, debts and derivatives.\\
{\it 2). Money markets:} They belong to the debt market. They trade in highly liquid and in a short-term debt like cash and foreign currency transactions, etc. For this reason they operate in a separate market.\\
{\it 3). Indexes:} They are defined as the weighted average of a basket of securities of a particular market. Although they are part of the capital market, they are treated different due to their importance and depth. 

\subsection{Definitions of Risk and Return}  

Any investor would like to maximize the return and minimize the risk. Among these two quantities, the risk is a complex quantity which involves the analysis of future uncertainties. We can define a fractional rate of return as 

\begin{equation}
R=\frac{S(t+T)+d-S(t)}{S(t)}.
\end{equation}               
Here $T$ is the time during which the stock is kept, and $d$ are the dividends worth. The risk on the other hand, is the degree of uncertainty for the stock price in the future. When a person or an organization is investing, different possible scenarios are possible. The investors could estimate the expected return by using the expression \cite{5}

\begin{equation}
\bar{R}=\sum_s p(s)R(s)\equiv E(X).
\end{equation}   
Here $p(s)$ is the probability for some possible future scenario to happen and $s$ is just a variable connected with the different possible scenarios. $X$ is a random variable connected the expectation value. We can then quantify the risk by defining it as the degree of deviation with respect to this expected value as follows

\begin{equation}
R=E(X)\pm\sigma.
\end{equation}  
Here $\sigma$ is the standard deviation (with the corresponding likelihood) as it is usually defined in probability theory. The standard deviation $\sigma$ is the square-root of the variance, which is defined as

\begin{equation}
\sigma^2=\sum_sp(s)\left(R(s)-E(X)\right)^2\equiv E[(R-E(X))^2].
\end{equation}  
Then the risk is quantified by $\sigma$. A large value of $\sigma$ corresponds to a large risk and viceversa. There are cases where the calculated value of $\sigma$ is infinite. In such situations, it is more convenient to use a different method for quantifying the risk \cite{5, New2}. On the other extreme, risk-free instruments like fixed deposits in a bank for example, have zero risk and then in such a case $\sigma_{risk-free}=0$. In addition, in these cases the instantaneous rate of return $r(t)$ for a risk-free instrument is defined as the spot interest rate, or overnight lending rate. Then when $\sigma=0$, the equivalence defined as: $Risk-free\; rate\; of\; return=spot\; interest\; rate=r$ is valid.\\   
The return is naturally higher when $\sigma>0$ and this encourages the investors to take the risk. We can then define the {\it risk premium} as the amount by which the rate of return on high-risk investment is above the risk-free rate. Given an expected annual rate of return $E(X)$, then the risk premium (RP) is defined as

\begin{equation}
RP=\frac{E(X)-r}{\sigma}.
\end{equation} 
Then an investor is willing to assume risk as far as there is a high {risk return} associated to the investment. This is connected with what is called the principle of {\it no arbitrage} in finance, which suggests that for getting a rate of return higher than the {\it risk-free rate}, then there must be a risk associated to it. 

\subsection{Definitions of no arbitrage, martingales and risk-neutral measure} 

The term {\it arbitrage} means to get guaranteed risk-free returns {\it above} the risk-less return that one can get from the money market. This is done by entering simultaneously into different financial transactions which not necessarily happens in the same market. Note that the benefits of the arbitrage can be observed only when the investors are big brokerage houses. Small traders cannot get any benefit from arbitrage because of the costs of transaction, which cancels any benefit from arbitrage in such cases. Note however, that if the market is in equilibrium, then there are no opportunities of arbitrage. At the same time, the arbitrage is the process through which the market can reach the efficient condition when all the traders have the same amount of information.\\
Arbitrage is connected with the {\it martingale process}, which corresponds to the process of evolving the discounted value of a financial instrument such that the price of the financial instrument is free of arbitrage \cite{5, 23, New4, New5}. A martingale process is an ideal concept, with respect to which we can compare the real evolution of the market. Real markets for example have an inherent risk such that there is a risk premium associated to them which works as a motivation for the investors. \\
The martingale condition, obeyed by a process in an efficient market, corresponds to the {risk-free} evolution of a security. The martingale process is defined by the conditional probability 

\begin{equation}
E[X_{n+1}\vert x_1, x_2,..., x_n]=x_n\;\;\;:\;\;\;martingale,
\end{equation}        
given a stochastic process. This conditions is just the expected probability of the random variable $X_{n+1}$, given already the values of all the previous stochastic variables $X_1, ..., X_n$. When we analyze a financial problem, the stochastic variables under consideration are the future prices of the stock $S_1, ..., S_{N+1}$ at the corresponding times $t_1, ..., t_{N+1}$. Note that the martingale condition, requires us to discount stock prices for comparison \cite{5}.\\
If we consider a financial market which is {\it complete} and where the condition of {\it no arbitrage} holds, then {\it the fundamental theorem of finance} states that these two conditions are equivalent to the evolution in agreement with the {\it martingale} condition for the financial instruments.\\
Once the discounting factor is fixed, for the same market, the evolution of the financial instruments obeys the martingale condition only for some specific (unique) risk-free probability \cite{New4, 42}. By considering that the risk-free evolution discounted stock price $e^{-int_0^tr(t')dt'}S(t)$ follows the martingale condition, it can be demonstrated that the present value of the stock is equal to the conditional probability of the discounted value of the equity at a future time $t$. Then we have

\begin{equation}
S(0)=E\left[e^{-\int_0^tr(t')dt'}S(t)\vert S(0)\right].
\end{equation}     
This result is general. This previous condition guarantees the absence of arbitrage opportunities for the derivatives instruments of the security. The opposite is also true, namely, the absence of arbitrage opportunities is related to the existence of a martingale condition. Such existence is what is called {\it The fundamental theorem of Finance}. 

\subsection{Hedging}

{\it Hedging} is the procedure of canceling the random fluctuations of an instrument with those of another one when they are included in a single portfolio. When the cancellation is perfect, then we say that we have a perfect {\it hedged} portfolio. Then in such a case, we have a zero risk portfolio. If we have two portfolios with the same return, then the one where the Hedging techniques are applied is by far superior since it has a lower risk associated to it \cite{5}. In practice usually is impossible to have a perfect {\it hedged} portfolio.\\
The process of {\it hedging} is not free and there is a cost of transaction associated to it. This is the price of reducing the risk. In some situations, {\it hedging} is undesirable because the investors would sacrifice future payoffs coming from "good" random fluctuations \cite{5}. Not all the fluctuations in the price of a financial instrument can be {\it hedged} and for the cases where we have a complete market, in principle it is possible to {\it hedge} every financial instrument. In practice, perfect {\it hedging} depends on the available instruments in the market. This is by the way, the main motivation for finding derivative instruments. \\
The process of {\it Hedging} requires at least a pair of instruments with correlated fluctuations. Such correlations are only possible if the instruments depend on each other. That's why this process involves a {\it derivative} instrument together with the associated primary one. Here we consider the example explained in \cite{5} and discussed here where we can consider the reduction of the risk of a security represented by a stochastic variable $S(t)$. The reduction of the fluctuations of this variable can be achieved by the introduction of a second instrument which is the derivative of $S(t)$, represented as $D(S)$ \cite{5}. For a perfect {\it Hedging}, we define the {\it hedged} portfolio as $\Pi(S, t)$. Following the example given in \cite{5}, we can define a composed portfolio as

\begin{equation}
\Pi(S, t)=D(S)-\Delta(S)S.
\end{equation} 
Here $\Delta (S)$ represents a short selling worth of the stock $S$ and the previous expression implies that the investor is buying on a single derivative $D(S)$. We can say that the portfolio $\Pi(S, t)$ is perfectly {\it hedged}, if there are no random fluctuations in its time evolution. Then the time-derivative

\begin{equation}
\frac{d\Pi(S, t)}{dt}=r(t)\Pi(S, t),
\end{equation}  
has not fluctuations associated to it if it is perfectly {\it hedged}. Then there is no risk in this portfolio. By using the principle of {\it no arbitrage}, then we conclude that the rate of return must be equal to the risk-free spot rate $r(t)$ and this explains why this quantity appears on the right-hand side of the previous equation. For closing this part, here we enumerate the necessary conditions for {\it hedging}, they are\\
1). The market must trade in the derivative instrument which is intended to be used for the process.\\
2). It is necessary to know the detail description of the fluctuations of the stock price $S(t)$, as well as the functional relation between $D(S)$ and $S$. Additionally, it is necessary to evaluate the existence of $\Delta (S)$ and its functional relation with $S$.\\
3). The market has to have enough liquidity, such that the process can be done continuously in time.  

\subsection{Further analysis on derivative securities}

The {\it derivative} instruments are obtained from the underlying financial instruments as could be perceived before. The classical example is the cash flow which is the derivative of the prices of the underlying instruments \cite{5}. The derivatives can be classified in three categories, they are {\it Forwards, Futures and Options}.

\subsubsection{Forward contracts}

The {\it forward} contract is the one done between a buyer and a seller, such that its initial value is zero. The payments are then agreed to be done in a future time $T$ at a price specified at the initial time $t$. The price is determined by the forward price $F(t, T)$ which reflects the actual price and the interest rates. The value of a forward contract fluctuates until its maturity defined at $T$. Then the value of the contract is 

\begin{equation}
S(T)-F(t, T),\;\;\;\;\;if\;long,
\end{equation}    

\begin{equation}
F(t, T)-S(T),\;\;\;\;\;if\;short.
\end{equation}
The expressions "long" is used for the buyer of the product and "short" is defined for the seller. The cash flow, at the maturity contract can be positive or negative depending on the price of the product at $T$. 

\subsubsection{Future contracts}

The future contracts, are similar to the forward contracts in the sense that there is an agreement between two parties for  the delivery of a product in a future time $T$. However, in this case there is an intermediate agent ({\it clearing house}) in the contract. The intermediate agent imposes margin payments of the buyer and the seller for reducing the owner party risk, as well as increasing liquidity and with the purpose of creating a series of cash flows from the starting time of the contract until its maturity. \\
At the starting of the contract, the value of the contract at the time $t$ is zero. However, it is assigned a notional {\it fair price} at the moment of writing the contract. If we define the {\it future contract} as $A(t, T)$, then this means that $A(t_0, T)\neq0$ for an initial time $t_0$.\\
For this kind of contracts, there is a daily maintenance margin if the price moves from its initial value. This daily maintenance receives the name of "marking to market". The value of the {\it future contracts} fluctuates with the random fluctuations of the asset. In addition, an initial margin is paid on initiating the contract \cite{5}.\\
When the contract matures, the value of the future contract converges to the value $S(T)$ due to the arbitrage. If $A(T, T)>S(T)$, then the seller can short the futures contract and then buy the stipulated amount from the market at the price $S(T)$. In this way a risk-less profit is done. On the other hand, if $A(T, T)<S(T)$, then the companies buying the product will arbitrage by entering into a long futures contract and then they will be able to get the asset below the market price. It is for this reason that at maturity, we have the condition

\begin{equation}
A(T, T)=S(T),
\end{equation}          
where in general, during the duration of the contract, $A(t, T)$ can be larger or smaller than the price of the security. In summary, {\it Future contracts} are more liquid than the {forward} contracts because they are standardized and being traded in the exchange market. The main difference between the two types of contracts is that for the case of {\it forward} contracts, there is only one-time cash flow for the period corresponding to maturity. On the other hand, for the case of {\it future} contracts, there is a series of cash flows which corresponds to the already-mentioned daily maintenance corresponding to the "marking to market" procedure. 

\subsection{Options}

An Option is defined as a derivative which can be written on any security, including other derivative instruments \cite{5}. More specifically, it is a contract for buying or selling. The contract can be entered by either, the buyer or the seller. For the cases of {\it forward} and {\it future} derivatives, the seller gets the compromise of delivering the corresponding asset once the contract is done. On the other hand, the investor does not care the asset but rather the profit which can be made from the contract.\\
We can give a few examples for different options. For the case of the European {\it call option}, the seller has the obligation to provide the stock at a given price (pre-determined) at some fixed moment in the future. The buyer on the other hand, does not have any obligation, he/she rather has the right to {\it exercise} or not the option. This decision is taken depending on whether or not the price of the stock on maturity is less or larger than the pre-determined price. In other words, an intelligent buyer, is able to take the following decisions

\begin{eqnarray}   \label{cr7}    
If\;\;\;S>K,\;\;\;buy,\nonumber\\
If\;\;\;S<K,\;\;\;Do\;not\;buy.
\end{eqnarray}
Here $S$ is the price of the stock on maturity and $K$ is the pre-determined price which the parties agreed before since the beginning of the contract. It is clear that the buyer makes a profit with the first condition illustrated in the previous equation. Different situation appears when we consider the case of the holder of a {\it put option}, where the power of decision is over the seller of the option \cite{5}. The form of the pre-determined price $K$ is called {\it payoff function} of the option \cite{Europe}. Further studies about Options can be found in \cite{Europe, Op2, Op3} and some quantitative methods can be explored in \cite{Optionagain, Option}.

\subsection{Path-independent options}

Some Options can be defined by the {\it payoff function} which only depends on the value of the underlying security at maturity. These are defined as path-independent options. The European options (both, the call and the put options) are examples of these kind of cases. 

\subsubsection{The European call option}

This can be represented as the function $\psi(S(t), t)$. At maturity, it is represented as \cite{5}

\begin{eqnarray}
g(S)=\psi(S(T), T)=S(T)-K,\;\;\;S(T)>K,\nonumber\\
g(S)=\psi(S(T), T)=0,\;\;\;S(T)<K.
\end{eqnarray}         
Here we have defined the pay-off function as $g(S)$. Whether the buyer buys or not the security is decided in agreement with the criteria (\ref{cr7}). 

\subsubsection{The European put option}

We can define it by $P(t)$ and its price is the same as in the case of the call option but with the difference that now the holder has the option of selling the security $S$ at the price $K$ \cite{5}. An interesting relation can be found between the {\it put} and the {\it call} option by using a {\it no arbitrage} argument \cite{Europe}

\begin{equation}
C(t)+Ke^{-r(T-t)}=P(t)+S(t), \;\;\;t\leq T.
\end{equation} 
This interesting relation is called the "Put-Call parity". 

\subsection{Path-dependent options}

As its name suggests, these Options correspond to the cases where the payoff functions depend on the path that the security takes before expiration \cite{5, Europe}. This is the case of the American option which can be exercised at any time before the expiration of the contract. This quality is what precisely makes it path-dependent, given the fact that the price of the security fluctuates in time. 
Everything else is similar to the European option. The {\it no arbitrage} arguments demonstrate that given a security which does not pay dividends, both, the American call option and the European call option have the same value. However, under the same conditions, the American {\it put option} has a higher value than the European put option.\\
The Asian option is another example of path-dependent option. In this case, the payoff function depends on the average value of the security during the whole duration period $T$. Other examples of path-dependent options can be seen in \cite{5}.

\subsection{Equations defining the dynamics of an option}

We can define the fundamental problem of option pricing as in \cite{5}. Given a payoff function $g(S)$, for an option maturing at the time $T$, we are interested in knowing the price of the option $\psi(S, t)$ before maturity $t<T$, given the instantaneous value of the security $S(t)$ at the same time. In this case, the final value of the option is defined by the known condition $g(S)=\psi(S, T)$. For the path-independent case, we then have a {\it final value} problem, trying to find the value of the option at earlier times $\psi (S, t)$. Since the value of the option depends on the value of the security, it is useful first to explore how the security changes. This problem is normally formulated by using the stochastic differential equation

\begin{equation}   \label{rainbowl}
\frac{dS(t)}{dt}=\phi S(t)+\sigma SR(t).
\end{equation}      
Here $\phi$ is the expected return of the security and $\sigma$ is the volatility. In addition, $R(t)$ is just a Gaussian white noise with zero mean. For the special case of zero randomness, we have the solution $S(t)=e^{\phi t}S(0)$. The white noise signal satisfies certain properties which can be seen in \cite{5}. 

\section{The Black-Scholes equation}   \label{BSEAMA}

In order to get the value of an option at some specific time, then the condition of {\it no arbitrage} must be satisfied. In fact, the options can be priced if we can make a perfect {\it hedge}. This observation is due to the Black and Scholes whose realized that we can have a certainty in the value of the option if there is no uncertainty associated to it. This observation seems to be logic. Under perfect {\it hedging}, then we have a {\it risk-free} rate of return given by the spot interest rate $r$ as has been explained before. A perfect hedging at all times implies the necessity of knowing the evolution of the option. The Black-Scholes idea suggests the creation of a hedged portfolio which is independent of the white noise $R$ \cite{Op2, Op4}. Following \cite{5}, we can define the portfolio

\begin{equation}      
\Pi=\psi-\frac{\partial \psi}{\partial S}S.
\end{equation}
This is a portfolio where an investor holds the option and then {\it short sells} the amount $\frac{\partial \psi}{\partial S}$ for the security $S$. By using the Ito calculus (stochastic calculus) \cite{5, hereag}, it is possible to demonstrate that

\begin{equation}   \label{BS}
\frac{d\Pi}{dt}=\frac{\partial \psi}{\partial t}+\frac{1}{2}\sigma^2S^2\frac{\partial^2 \psi}{\partial S^2}.
\end{equation}
Here the change in the value of $\Pi$ is deterministic \cite{5}. The random term has disappeared due to the choice of portfolio. Since here we have a risk-free rate of return for this case (no arbitrage) \cite{Europe, Epjons}, then the following equation is satisfied

\begin{equation}
\frac{d\Pi}{dt}=r\Pi.
\end{equation} 
If we use the result (\ref{BS}), then we get

\begin{equation}   \label{BSeq}
\frac{\partial \psi}{\partial t}+rS\frac{\partial \psi}{\partial S}+\frac{1}{2}\sigma^2S^2\frac{\partial^2\psi}{\partial S^2}=r\psi.
\end{equation}
This is the Black-Scholes equation \cite{Op2, Op3, Merton2}. There are some points to remark about the Black-Scholes equation. The first point to remark is that this equation does not depend on the expectations of the investors. Note that the parameter defining this is $\phi$, which appears in eq. (\ref{rainbowl}). In other words, in the Black-Scholes equation, the security (derivative) price is based on a risk-free process. This equation is indeed ideal and it is based on the ideal condition of a (risk-less) hedged portfolio. Extensions have been analyzed in \cite{Optionagain, Extensions}. Other possibilities for analyzing the option pricing can be found in \cite{Epjons, Againla?}. The Basic Assumptions of the Black-Scholes equation are:\\
1). The spot interest rate $r$ is constant.\\
2). In order to create the hedged portfolio $\Pi$, the stock is infinitely divisible, and in addition it is possible to short sell the stock.\\
3). The portfolio satisfies the no-arbitrage condition.\\
4). The portfolio $\Pi$ can be re-balanced continuously.\\
5). There is no fee for transaction.\\ 
6). The stock price has a continuous evolution. \\
One comment related to this last point is that for pure stochastic processes, the Black-Scholes equation cannot be applied. The reason is that a perfect hedging is not possible in such cases \cite{Merton2, BHR}. These previous points, remark that the Black-Scholes equation is an idealization of the stock market. However, the starting point for every analysis in any science is to develop and understand the ideal case and then based on it, real situations can be analyzed by doing the corresponding extensions. In this paper we do not intend to make any change in the Black-Scholes equation. We will rather re-express it in a convenient form in order to introduce a new Mathematical tool for its analysis.

\subsection{The Black-Scholes equation expressed as a Schr\"odinger equation}   \label{The example}

We will explain how the eq. (\ref{BSeq}), can be expressed as an eigenvalue problem after a change of variable. The resulting equation will be the Schr\"odinger equation with a non-Hermitian Hamiltonian as we will demonstrate in a moment. For eq. (\ref{BSeq}) consider the change of variable $S=e^x$, where $-\infty<x<\infty$. Under this change, then the equation becomes

\begin{equation}   \label{BSHamiltonian2}
\frac{\partial\psi}{\partial t}=\hat{H}_{BS}\psi,
\end{equation} 
with 

\begin{equation}   \label{BSHamiltonian}
\hat{H}_{BS}=-\frac{\sigma^2}{2}\frac{\partial^2}{\partial x^2}+\left(\frac{1}{2}\sigma^2-r\right)\frac{\partial}{\partial x}+r.
\end{equation}
Note that the resulting Hamiltonian is non-Hermitian since $\hat{H}\neq\hat{H}^+$ \cite{5}. Once again here we remark that the fact that the BS equation can be expressed in the form (\ref{BSHamiltonian}), does not mean the the financial market is quantum. This only means that some equations in Finance, can be expressed as an eigenvalue problem and then take the form of the Schr\"odinger equation in Quantum Mechanics. In this way, we can use the same Mathematical techniques of Quantum Mechanics for doing some interesting analysis. This is a significant advantage because the techniques employed in Quantum theory are widely explored and well understood in some cases. Note that since the Hamiltonian represented in eq. (\ref{BSHamiltonian}) is non-Hermitian, then the evolution in time of the Option is non-unitary in general (in addition, the Hamiltonian non-necessarily obeys the $PT$ symmetry). This means that the probability is not necessarily preserved in time, although it is certainly well-defined and its total value is equal to one. As a basic example about this statement, imagine we have $50$ coins (with different values) in a bag and we want to calculate the probabilities of getting specific values of coins. The values for the probability will be naturally $P=1/50$ at such instant of time. Now imagine that one minute later $10$ more coins are included inside the bag. The new probability for extracting the coins will be $P=1/60$. As you can see, at both instants of time, the sum of the probabilities is equal to one. However, since new information was input for the second instant, naturally the probability is not conserved due to the change of the sample space. A more general explanation of this situations comes out if we define the un-normalize the wave-function as $\psi^*\psi=A_1^2$ at one instant of time $t_1$ but then in a subsequent instant $t_2$ we have $\psi^*\psi=A_2^2$ with $A_1\neq A_2$. At both instants, the total probability is one (this is precisely the purpose of introducing the normalization factor) if we define the normalized function as $\psi/\sqrt{A_i}$ ($i=1,2$). However we cannot compare directly the two instants without considering the flow of information through the boundaries of the system when we go from one instant to the other. In other words, the normalization factor changes in time. In general, there are some cases in ordinary Quantum Mechanics, as well as in Quantum Field Theory, where it is interesting to explore non-Hermitian Hamiltonians (Lagrangians) \cite{Ellis, Ellis2}. Based on the previous explanations, we cannot expect the financial market to obey unitarity. The reason for this is simply because the market is not a closed system and there are many external factors influencing its behavior as for example it is the amount of people and organizations trading at some instant of time. Then the assumption of unitarity makes no sense at all and the Hamiltonian must be non-Hermitian. We will talk more about this important statement later in this paper.  

\section{The standard double-slit in Quantum Mechanics: Basic concepts}   \label{II}

\begin{figure}
	\centering
		\includegraphics[width=10cm, height=8cm]{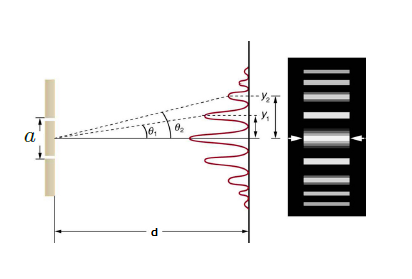}
	\caption{ The classical version of the double slit experiment showing the interference pattern of
impacts over the screen. Taken from \cite{7}}
	\label{Fig1}
\end{figure}

The previous two sections introduced some notions and basic concepts of financial markets. We also explained how to obtain the Black-Scholes equation and how to express it in the Hamiltonian form in eqns. (\ref{BSHamiltonian2}) and (\ref{BSHamiltonian}). Note that we have never changed the equation, all what we did was to write it in a different and convenient form. After expressing the BS equation in this way, we can use over it the same tools used in Quantum Mechanics. Then in this section and the next, we will explain two important concepts taken from ordinary Quantum Mechanics, namely, the standard double-slit experiment and its weak-value representation. Subsequently we will use the same techniques explained in this and the next section on the Black-Scholes Hamiltonian. Here we explore the basic formalism in order to understand the standard results of the double slit experiment in Quantum Mechanics \cite{6}. The results of the double slit experiment
are independent on the nature of the particles under consideration, then we expect them to
model some specific aspects in the behavior of the Stock Market. We start by saying that a
particle moving freely over the space obeys the following equation             

\begin{equation}
i\hbar\frac{\partial}{\partial t}\psi({\bf r}, t)=\hat{H}\psi({\bf r}, t).
\end{equation}   
Here $\hat{H}$ is the Hamiltonian of the particle and $\hbar=h/2\pi$ is the reduced Planck scale. Note
that for the case where the particle under analysis is moving freely, the Hamiltonian only
contains the kinetic energy contribution defined as

\begin{equation}
\hat{H}=\frac{\hat{{\bf p}}^2}{2m}.
\end{equation}
The action of the Hamiltonian over the wave function $\psi({\bf r}, t)=<{\bf r}\vert\psi>$, which is defined over the space ${\bf r}$ is obtained by understanding the way how the momentum operator behaves under the same conditions. The action of the momentum operator over the position space is given by

\begin{equation}   \label{Thisone}
\hat{ p}\psi(r,t)=-i\hbar\frac{\partial}{\partial r}\psi(r,t).
\end{equation}
Here we are working under the assumption of spherical symmetry. Then the angular derivatives are irrelevant. In general, in three dimensions, the momentum operator is defined as

\begin{equation}
\hat{{\bf p}}\psi({\bf r}, t)=-i\hbar\nabla_{\bf r}\psi({\bf r}, t).
\end{equation}
If we replace the previous result in eq. (\ref{Thisone}), then we get

\begin{equation}
i\hbar\frac{\partial}{\partial t}\psi(r, t)=-\frac{\hbar^2}{2m}\nabla^2\psi(r,t).
\end{equation}
Note that for a free-particle we take the potential term as $\hat{V}(\hat{r},t)=0$, namely, the potential
term vanishes for this case. It is clear that the solution of the previous equation is simply
a superposition of plane waves. More complicated solutions appear when we consider a
potential term in addition to the kinetic one. In the case of a double slit experiment, the
particle behaves as evolving freely (zero potential), except for the fact that it will interact
with a potential barrier which is infinity at some specific time of the evolution of the system,
except at the two locations (points) corresponding to the slits which are separated by a
distance $a$. The particle can only penetrate through these pair of points. The
figure (\ref{Fig1}), illustrates the pattern of impacts over a screen, when a large number of particles are able to cross the double slit. Interestingly, the pattern
of impacts reflects the dual character of a particle in Quantum Mechanics \cite{6}. Any particle,
which is able to penetrate through the slits, will have a wave-like behavior as far as the relation
$a<<\lambda=h/p$ is satisfied. Here $h$ is the Planck constant. The wave-like behavior of the
particle almost disappears for the case where we consider short wavelengths. In such a case,
the particle-like behavior dominates. This happens when $a>>\lambda$ and then the pattern of impacts over the screen corresponds to the expected classical behavior. After crossing the double slit, the
particle will impact the screen at an angle with respect to the horizontal line, defined as $\theta$.
Then the pattern of the maximal of intensity fringes over the screen, obeys the relation

\begin{equation}
\Delta y=d\theta\approx n\lambda,
\end{equation}
taking into account the approximation $sin\theta\approx\theta$ valid for small angles. This approximation is correct as far as the distance $d$ between the screen and the plate where the two slits are located is very large, namely, when $d>>a$. Note that $\Delta y$ corresponds to the vertical distance
between the different maximal of intensity over the screen. Note also that when $\lambda\to0$, then
$\Delta y\to0$ for any particle crossing the two slits and then the classical behavior is recovered
as expected.

\section{THE WEAK-VALUE VERSION OF THE DOUBLE SLIT EXPERIMENT}   \label{III}

A deeper understanding of the double-slit experiment is accomplished when we consider
the weak-value formulation. The weak-value is {\bf a complex quantity} defined as \cite{1}

\begin{equation}   \label{7}
A_w=\frac{<\psi\vert A\vert\phi>}{<\psi\vert\phi>}.
\end{equation}
Here $\vert\phi>$ corresponds to the pre-selected state and $\vert\psi>$ corresponds to the post-selected one. Note that the weak-value is not an eigenvalue; then even if the operator
$\hat{A}$ is Hermitian, the weak-value is a complex number which has a magnitude and phase associated to it. We will apply this concept to the double-slit experiment in Quantum Mechanics. The time evolution
for the weak-trajectory can be described by using the following expression
	
\begin{equation}   \label{8}
A_w(t)=\frac{<\psi\vert U(T-t)AU(t)\vert\phi>}{<\psi\vert U(T)\vert\phi>}.
\end{equation}	
This expression represents the outcome of a weak-measurement at a time $t$ inside the interval
$[0,T]$. In this case, we pre-select a state defined at $t=0$ by $\phi$ and then we post-select a
state defined at $t=T$ by $\psi$. The evolution of the system is described through the operator $U(t)=e^{−iHt/\hbar}$, which is unitary when $\hat{H}$ is Hermitian. In such a case, we claim that the evolution is unitary for the quantum state defined by the Hamiltonian $\hat{H}$ of the system under consideration. The previous equation represents the weak-value measurement
of any Hermitian operator, for a pre-selected state $\vert\phi>$ evolving forward in time and a
post-selected state $\vert\psi>$ evolving backward in time. In order to get a finite value for this
quantity, there must be an overlap between the pre-selected state and the post-selected one.
If the operator $\hat{H}$ is non-Hermitian, the evolution of the system is not necessarily unitary and some modifications will appear in the calculations unless the $PT$ symmetry is satisfied in the system \cite{Ellis, Ellis2}. For the case of the double slit experiment, we will take the pre-selected state as

\begin{equation}   \label{9}
\vert\phi>=\frac{1}{\sqrt{2}}\left(\vert x_i>+\vert-x_i>\right).
\end{equation}
Here $x_i$ corresponds to the location of the upper slit with respect to the center and $-x_i$
corresponds to the location of the lower slit with respect to the center. Note that that the
separation between the slits is here taken as $2x_i=a$. The Post-selected state can be any
point over the screen and here we will consider it as

\begin{equation}   \label{10}
\vert\psi>=\vert x_f>.
\end{equation}
If we consider a free-particle, it is well-known that the Feynman Kernel for the transition
amplitude becomes 

\begin{equation}   \label{11}
<x_f\vert U(T)\vert x_i>=\left(\frac{m}{2\pi i\hbar T}\right)^{1/2}exp\left(i\frac{m(x_f-x_i)^2}{2\hbar T}\right).
\end{equation}
Based on this information, we can then find easily that the pattern over the screen obeys
the result

\begin{equation}
\vert <\psi\vert U(T)\vert\phi>\vert^2=1+cos\left(\frac{2m x_ix_f}{\hbar T}\right).
\end{equation}
we can obtain the weak-trajectory by using eq. (\ref{8}) but applied for the position operator,
getting then the result 

\begin{equation}   \label{13}
x_w(t)=\frac{<x_f\vert U(T-t)xU(t)\vert\phi>}{<x_f\vert U(T)\vert\phi>}.
\end{equation}
If we use the result suggesting that the weak-trajectories follow the classical ones \cite{4}, then
eq. (\ref{13}) becomes equivalent to

\begin{equation}   \label{14}
x_w(t)=x_f\frac{t}{T}-ix_i\left(1-\frac{t}{T}\right)tan\left(\frac{m x_ix_f}{\hbar T}\right).
\end{equation}
This describes the weak-trajectory of the particle over a complex plane. It can be demonstrated that when the interference is constructive, the imaginary part vanishes. On the other
hand, for destructive interference, the imaginary part diverges \cite{4}. The real part just follows
the classical trajectory which is an average of all the possible paths from $x_i$ to $x_f$.

\section{THE DOUBLE SLIT EXPERIMENT IN QUANTUM FINANCE: THE WEAK-VALUE APPROACH}   \label{IV}

We will develop the concept of Weak-Value in Quantum Finance. As an example, we will consider the Black-Scholes equation expressed as an eigenvalue problem as in eq. (\ref{BSHamiltonian2}). Note that here we do not make any change over the original Black-Scholes equation. Instead we have expressed the same equation in a convenient form such that we can apply the same Mathematical techniques of Quantum Mechanics. The resulting Hamiltonian $\hat{H}_{BS}$ is non-Hermitian and this means that we should not expect conservation of probability. There is no reason neither for expecting the $PT$ symmetry to be satisfied by the Hamiltonian. As we have remarked previously, we cannot expect unitarity in the evolution of an Option because there is a permanent inflow and outflow of information over the market. Then the non-unitarity is an expected phenomena in this case and the pattern of "Impacts" over the "Screen" (The pattern of prices which we are able to read for a given final time), is different with respect to the ordinary Quantum-Mechanical case. In Quantum finance, the impact over the screens represents the final price for the Option. The double-slit
on the other hand, will be taken as our pre-selected condition. This gives us two
initial Option prices differing by a distance defined by the separation between the slits. The potential term is taken to vanish everywhere,
except for the fact that we have a double-slit at some specific instant. At the time related to
the location the two slits, the option becomes worthless, except over the two points where
the slits are located. Then we can for example, pre-select the quantum state exactly in
the same way as has been done in eq. (\ref{9}). The Post-selected state will be equivalent to
the same state illustrated in eq. (\ref{10}). What will make the difference at the end, will be
the definition of the Kernel representing the amplitude for a particle to propagate from one
point to another one as it was defined in eq. (\ref{11}). For the case of the financial approach
this corresponds to the pricing Kernel \cite{5}. This Kernel will depend on which Hamiltonian we
want to use and it will represent the amplitude for the evolution of the Option prices in going
from one initial value toward a final one. In what follows we will analyze the evolution of the
weak-value in Quantum finance by using the Black-Scholes Hamiltonian as an example and
then deriving a generic result. Note that the method employed here is general and independent on which Financial equation is used as far as it can be expressed as a Scr\"odinger equation (eigenvalue equation). If we use the standard approach of Quantum finance, we have
to take into account that usually the formulation is developed by considering that the time $t$
is defined implicitly through the function $\tau=T-t$ for an interval defined as $t\;\;\epsilon\;\;[0,T]$. This
means that $\tau$ represents the evolution backwards in time. Then at the moment of defining
the weak-value in Quantum Finance, we will make an inversion on the roles taken by the
pre-selected and the post-selected states. As a consequence of this, we will take the final price for the Option $x_f$ as the pre-selected state and the initial state $\vert\phi>$ as the post-selected one from the financial perspective.
 
\subsection{Comments on non-Hermitian Hamiltonians and their natural appearance in Open systems}

The Financial market is an open system. As a consequence of this, the Hamiltonian has to be non-Hermitian \cite{Newagain, Newagain2}. In an open system, it is the interaction between the system and the environment (flow of information) what reproduces the effect of loss of information which is effectively modeled by non-Hermitian Hamiltonians. In fact, some authors claim that real Quantum systems are always open and there is always some degree of non-unitarity and then the full expansion of the Hamiltonian has to be non-Hermitian. In some simplified situations, the Hamiltonian can be separated in Hermitian plus non-Hermitian contributions in the following form \cite{Newagain}

\begin{equation}   \label{simpledec}
\hat{H}^{eff}=\hat{H}_0+\sum_cV_{0c}\frac{1}{E^+-\hat{H}_c}V_{c0}.
\end{equation} 
The closed system is naturally described by the Hermitian contribution $\hat{H}_0$. $(E^+-\hat{H}_c)^{-1}$ is the Green function and the Hamiltonian $\hat{H}_c$ describes the environment of decay channels. $V_{0c}$ and $V_{c0}$ represent the couplings of the closed system (Hermitian) with the different channels (paths connecting with) of the environment \cite{Newagain}. Then whenever there is an interaction between a system and the environment, the full Hamiltonian describing the system has to be non-Hermitian. This implies non-unitarity or non-conservation of the probability in the sense that the normalization factor which makes the sum of probabilities equal to one, changes in time. As a basic example about this issue has been mentioned in Sec. (\ref{The example}). Another important point about Open systems, is the fact that in some situations, they have associated to them an arrow of time \cite{Newagain3}. This means that whenever a system is open, not only its Hamiltonian is non-Hermitian, but in addition, there exists the possibility of violating the time-reversal symmetry explicitly. This important statement has been demonstrated before in \cite{Newagain3} and it is a natural consequence of the fact that non-Hermitian Hamiltonians have complex eigenvalues. The possible violation of the time-reversal symmetry can be explained easily if we consider again the eigenvalue equation

\begin{equation}   \label{Ham111}
\hat{H}\vert\psi>=E\vert\psi>.
\end{equation}
It is already known that the time-reversal is represented by an antiunitary operator of the form $T=KU$, where $U$ is a unitary operator and $K$ represents complex conjugation \cite{Peskin}. By applying the time-reversal operator to the previous equation, we obtain

\begin{equation}   \label{Ham112}
T\hat{H}\vert\psi>=TE\vert\psi>=E^*(T\vert\psi>)=\hat{H}(T\vert\psi>).
\end{equation}
With a complex eigenvalue $E^*$, as it is the case of the anti-Hermitian Hamiltonians, then eqns. (\ref{Ham111}) and (\ref{Ham112}) represent different eigenvalues and  then the eigenstates might break the time-reversal symmetry in general. This is typical in situations where there is a permanent flow of information through the boundaries of the system. Note however that in \cite{Newagain}, it is remarked that there might be cases where the time-reversal symmetry is not necessarily broken in open systems, even under the non-Hermiticity of the Hamiltonian. In the case of the Stock market however, the processes are highly irreversible and we should expect the development of an arrow of time or increasing in the entropy \cite{NA4}. Here we will not analyze this issue in detail but what is important to remark is that non-Hermitian Hamiltonians are very common in physics as well as in other areas and that Open systems must be represented in this way. Finally in this subsection we want to remark that the complex eigenvalues corresponding to different eigenstates can meet in the complex plane under special circumstances \cite{Newagain}; reproducing in this way interesting topological effects \cite{kato}. This is related to the definitions of exceptional points, where the conditions of orthogonality are not necessarily satisfied. For example, if we define the quantity \cite{Newagain}

\begin{equation}
r_k\equiv\frac{<\Phi_k^*\vert\Phi_k>}{<\Phi_k\vert\Phi_k>},
\end{equation}
then this quantity would vanish very near the exceptional point, while it becomes unity far away from them. Fortunately, our weak-value formulation is simpler and does not have to deal with this issue, although it is important to understand that there are different kind of approaches and analysis related to the use of non-Hermitian Hamiltonians representing open quantum systems. In the case of the financial market for example, the simple decomposition defined in eq. (\ref{simpledec}) is possible. In fact, the Black-Scholes Hamiltonian (\ref{BSHamiltonian}) can be decomposed as

\begin{equation}
\hat{H}_{BS}=H_{BS}^{H}+\hat{H}_{BS}^{AH},
\end{equation}     
where the subindices $H$ and $AH$ correspond to the cases of Hermitian and Anti-Hermitian portions of the Hamiltonian. It is possible to demonstrate that these portions are defined as \cite{5}

\begin{equation}
H_{BS}^{H}=-\frac{\sigma^2}{2}\frac{\partial^2}{\partial x^2}+r,\;\;\;\;\;\;\;\;\hat{H}_{BS}^{AH}=\left(\frac{1}{2}\sigma^2-r\right)\frac{\partial}{\partial x}.
\end{equation}  
We conclude then that the non-Hermiticity of the Black-Scholes Hamiltonian is due to the drift term (velocity potential term) and this is related to the fact that the Financial market is open with permanent flow of information and the changes in the amount of traders is related to this flow of information. The information corresponds to sudden changes in the politics of a country among other facts. In order to finish this small subsection, we want to remark once again that non-Hermitian Hamiltonians describe realistic situations because the systems are never completely isolated. Isolation is only an idealization or approximation to the reality. One typical example happens in condensed maatter physics where interesting physical effects related to the phenomena of superfluidity appear. In \cite{NA5}, it was demonstrated that unconventional phase transitions might occur when non-Hermitian BCS Hamiltonians are used for describing ultracold atoms. There exists experimental evidence about this and the details can be found in \cite{NA5}.   
        
\subsection{Weak-value for operators evolving in agreement with non-Hermitian Hamiltonians}

Non-Hermitian Hamiltonians describe non-unitary evolutions in general \cite{8}. Then their eigenvalues are in general complex numbers. Then for example, the action of the operator $U(t)=e^{−i\hat{H}t}$ over a ket can be represented as

\begin{equation}
\vert z_n(t)>=e^{-i\hat{H}t}\vert z(0)>=e^{-i\omega_nt}e^{-\alpha_nt}\vert z_n>.
\end{equation}
Note that the evolution of the previous ket, corresponds to a damping oscillation. It is also possible to get an unbound oscillation depending on the sign taken by $\alpha_n$.
There might be cases where the wavelength is infinity, losing then completely the wave-like behavior. This happens for example if $\omega_n\to0$ for a finite value of
$\alpha_n$. One important consequence of the non-unitary evolution described by non-Hermitian Hamiltonians is the fact that observables which initially are compatible (zero commutator),
become non-compatible after some period of time (non-vanishing commutator). The opposite behavior can also happen, namely, a pair of non-commuting operators, can eventually commute after some time evolution \cite{8}. As we have remarked before, in some situations, it is possible to recover the unitarity condition for Non-Hermitian operators if the $PT$ symmetry is satisfied \cite{Ellis, Ellis2}.  

\subsection{The concept of weak-value applied to the Black-Scholes Hamiltonian}

Here we develop basically the same calculations of the previous section, but now taking into account that the pricing Kernel is defined as \cite{5} 

\begin{equation}   \label{16}
p_{BS}(x, \tau; x')=<x\vert e^{-\tau H_{BS}}\vert x'>, 
\end{equation}
with the Black-Scholes Hamiltonian appearing in this expression. There are some differences with respect to the definition of the ordinary Kernel defined
in eq. (\ref{11}). First, the time $\tau$ is not the ordinary time running forward. Instead, it is
defined as $\tau=T-t$, namely, as a variable running backward. The other difference is that
the exponential term in eq. (\ref{16}) is not imaginary as in the standard Quantum Mechanical
formulation. This is a natural consequence of the non-unitarity. Considering the Hamiltonian (\ref{BSHamiltonian}), then the pricing Kernel previously defined becomes

\begin{equation}   \label{18}
p_{BS}(x, \tau; x')=e^{-r\tau}\frac{1}{\sqrt{2\pi\tau\sigma^2}}e^{-\frac{1}{2\tau\sigma^2}(x-x'+\tau(r-\sigma^2/2))^2}.
\end{equation}
This is a normal distribution with mean equivalent to $x(t)+\tau(r-\sigma^2/2)$. Since $\tau$ defines a
time running backward, then we will re-define the weak-value in eq. (\ref{7}) as 

\begin{equation}
A_w^F=\frac{<\phi\vert A\vert \psi>}{<\phi\vert\psi>},
\end{equation}
exchanging then the roles of the pre-selected and post-selected state. The superindex means
"Financial" in order to distinguish this previous definition from the one usually evaluated
in ordinary Quantum Mechanics. Since what we want for this case is the evaluation of the
prices for the weak-value Options, then we have to work-out the following equation

\begin{equation}  \label{20}
x_w^F(t)=\frac{<\phi\vert U(T-\tau)x^FU(\tau)\vert x_f>}{<\phi\vert U(T)\vert x_f>}.
\end{equation}
This corresponds to the weak-trajectory for a the option price evolving from the final state
$\vert x_f>$ toward the two possible initial prices represented by the initial state $\vert\phi>$. The previous expression can be expanded as
follows

\begin{equation}
x_w^F(t)=\frac{<x_i\vert U(T-\tau)x^FU(\tau)\vert x_f>+<-x_i\vert U(T-\tau)x^FU(\tau)\vert x_f>}{<x_i\vert U(T)\vert x_f>+<-x_i\vert U(T)\vert x_f>}.
\end{equation}
By using the definition

\begin{equation}
x_w^\pm=\frac{<\pm x_i\vert U(T-\tau)x^FU(\tau)\vert x_f>}{<\pm x_i\vert U(T)\vert x_f>},
\end{equation}
then we get

\begin{equation}   \label{23}
x_w^F(t)=\frac{<x_i\vert U(T)\vert x_f>x^+_w+<-x_i\vert U(T)\vert x_f>x_w^-}{<x_i\vert U(T)\vert x_f>+<-x_i\vert U(T)\vert x_f>}.
\end{equation}
Note that here $U(t)=e^{−i\omega_nt\pm \alpha_nt}$ represents a non-unitary operator. For the case of ordinary Quantum Mechanics, $\alpha_n=0$ and then we recover explicitly the unitary evolution of the system, preserving in such a case the commutators over time. In order to evaluate the previous expression, we need to
know the value of the amplitude

\begin{equation}
<\pm x_i\vert U(T)\vert x_f>.
\end{equation}
This amplitude is already defined in eq. (\ref{18}) if we consider $\tau=T$, $x'=x_f$ and $x=x_i$.
Then the Kernel amplitude becomes

\begin{equation}   \label{25}
<\pm x_i\vert U(T)\vert x_f>=p_{BS}(x_i, T; x_f)=e^{-rT}\frac{1}{2\pi T\sigma^2}e^{-\frac{1}{2T\sigma^2}(\pm x_i-x_f+T(r-\sigma^2/2))^2}.
\end{equation}
If we evaluate the result (\ref{23}) by using the amplitude defined in eq. (\ref{25}), then we obtain

\begin{equation}   \label{26}
x_w^F(t)=x_f\left(1-\frac{\tau}{T}\right)+x_i\frac{\tau}{T}tanh\left(\frac{x_i}{T\sigma^2}\left(x_f-T\left(r-\frac{\sigma^2}{2}\right)\right)\right)
\end{equation}
Here we have used the relations

\begin{equation}   \label{27}
x_w^\pm(t)=(\pm x_i-x_f)\frac{\tau}{T}+x_f.
\end{equation}
It is important to notice that the pre-selected state $\phi$ considers the two possible prices $\pm x_i$. We have to remark that this is just a matter of Mathematical convenience. However, the prices for the Options are defined as positive. Then for example, we could take $-x_i=0$ and $x_i=a$ such that the condition $2x_i=a$ is satisfied. In such a case, the minimal price considered at the initial time $t=0$ is zero and $x_i$ would represent the highest price at that time. For the same case $x_f$ would define the final price of the Option. However, if $-x_i=c$, with $c$ representing some minimal price, then we have $x_i=a+c$ representing the larges possible price. For this second case, the final price has to be shifted correspondingly as $x_f^{real}=c+x_f$, with $x_f^{real}$ representing the real final observed price. Then it is important to take into account that it is Mathematically convenient to use the lowest and the largest possible prices at a moment of time as $\pm x_i$ with the corresponding final prices represented by $x_f$. The variables are defined with respect to some reference price $c$. Then we have to translate these variables to the corresponding positive numbers representing the prices of the Options by doing the appropriate shifts. In summary, $x_f$ is the final price measured with respect to $-x_i$, and $x_i$ is the largest possible initial price also measured with respect to $-x_i$, meanwhile $-x_i$ is the minimal possible initial price. The Figure (\ref{Fig2}) illustrates this situation. We can now test the result (\ref{26}). For example when $\tau=0$, we have $t=T$ and then the previous result is $x^F_w (T)=x_f$ as it should be. On the other hand, if we impose the condition $\tau=T$, which is equivalent to $t=0$, we get

\begin{equation}   \label{28}   
x_w^F(0)=x_itanh\left(\frac{x_i}{T\sigma^2}\left(x_f-T\left(r-\frac{\sigma^2}{2}\right)\right)\right).
\end{equation}
This result corresponds to the location of the weak-trajectory of the price for the Option at the initial
time. Note that since the initial price of the option is uncertain, due to the fact that we have
two possible results ($\pm x_i$), then the quantity in eq. (\ref{28}) will take a value between $x_i$ and $-x_i$ as can be easily verified. Whether the weak-value representing the price of the Option is closer to $x_i$ or closer to $-x_i$, depends on the value taken by the argument of the hyperbolic tangent function in eq. (\ref{28}). We will define the price in this equation as the "Weak"-price of the {option. The extension of the results for the case where we define $n$-prices for a stock at a given instant of time is
direct. It is also possible to extend this case to situations where we have multiple prices at one instant of time (non-necessarily equidistant), connected with multiple prices at a final time (not necessarily equidistant). The Figure (\ref{Fig3}) illustrates this situation, which can be used as a powerful predicting tool for circumstances where we have only partial information, after considering more realistic models. Realistic models can be done for example if we introduce some potential terms in the Hamiltonian as $H=H_{BS}+V$.

\subsubsection{\bf Applications of the double-slit approach in situations where we only have one price at a given instant of time}

Normally an Option can only take one price at a given instant. However, even in such situations, the concept of double slit is useful for two possible analysis. They are:\\

{\bf 1)}. For the cases where we want to know at a given instant of time, the degree of uncertainty of the price of the Option. In such a case, the weak-value for the price of the Option, represented in general by the quantity (\ref{20}) (red line in the Figure (\ref{Fig4})), is the expected price with the uncertainty bars (blue lines in Fig. (\ref{Fig4})) representing the separation between the slits. In these situations, the weak-trajectory for the price of the Option is defined by eq. (\ref{26}).\\ 
{\bf 2)}. For the cases where there is a large fluctuation for the price of an option in some interval of time as it is illustrated in Figure (\ref{Fig2}). The largest value of the price $O_1$ appears at a time earlier than the smallest price $O_2$ for the same Option. By considering an average over the interval, we can convert this problem to a double-slit experiment with $O_1=x_i$ and $O_2=-x_i$. The shorter is the interval under consideration, then we have a better approximation for the double slit experiment for this case. In other words, when there is a large fluctuation in the price, up to some order, the approximation of the double slit is valid after replacing the interval under consideration for a given instant defined by the average time of the same interval. By repeating this process multiple times through different intervals, we obtain a new graphic with the weak-value of the price representing the price of the Option with an uncertainty $\Delta x=2x_i=a$, which is equivalent to the separation between two slits. The new time interval under consideration then becomes shorter because it corresponds to a set of averaged intervals. Then this is a special case of the same situation analyzed in the item {\bf 1).} and then after averaging over the set of intervals, we obtain a situation also represented by the Figure (\ref{Fig4}). The approximation of the double slit can also be done in the opposite sense, namely, for the cases where $O_1<O_2$. In other words, the order of the prices defined in time does not change this analysis. \\\\
 
\begin{figure}
	\centering
		\includegraphics[width=10cm, height=8cm]{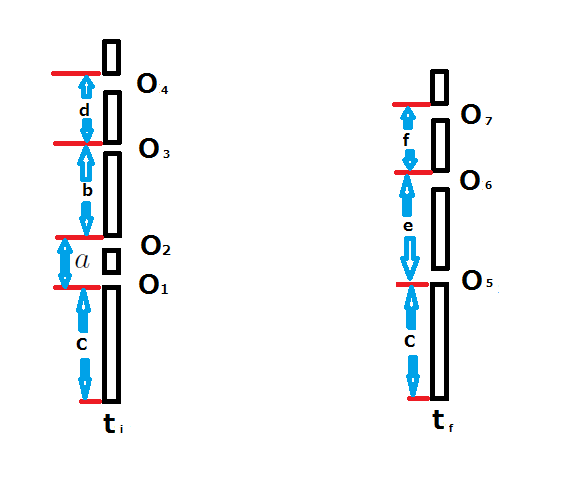}
	\caption{Extensions of the double slit case to the situations with multiple possible prices at an initial time, connected with multiple possible prices for the Options at a final time. Note that the difference between prices, namely, $O_n-O_{n+1}$ (with $n$ representing any integer number) is arbitrary. This case is important for situations where we have some partial information about the system and we want to make some predictions. The Weak value can then give us the expected trajectory for the price of the Option, based on the partial information provided.}
	\label{Fig3}
\end{figure}

\subsection{The generic result}

The result just obtained in this previous section, is generic in the sense that for other
Hamiltonians the result will be in essence the same. In general, given a Kernel amplitude
defined as

\begin{equation}
<\pm x_i\vert U(T)\vert x_f>=Ae^{f(x_i, x_f)},
\end{equation}
with $f(x_i,x_f)$ defined as a function of the initial and final price for the derivative, then the weak-trajectory of the price, will be defined as

\begin{equation}   \label{2Amazing}
x_w^F(t)=x_f\left(1-\frac{\tau}{T}\right)+x_i\frac{\tau}{T}tanh(f_{odd}(x_i, x_f)). 
\end{equation}
Here $f_{odd}(x_i, x_f)$ represents the the portion of the function $f(x_i, x_f)$ which is is odd
with respect to the initial stock price $x_i$. In fact, this function satisfies the condition
$f_{odd}(\pm x_i, x_f)=\pm f_{odd}(x_i, x_f)$.

\begin{figure}
	\centering
		\includegraphics[width=10cm, height=8cm]{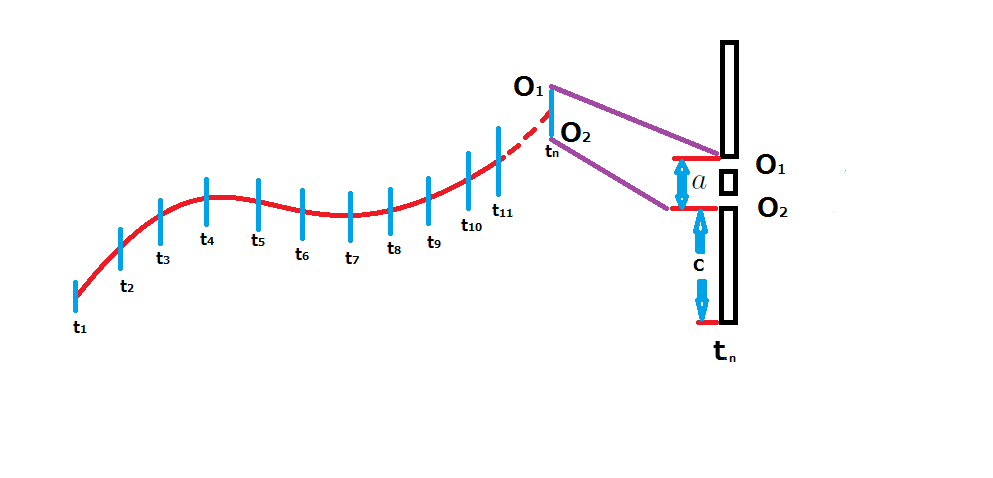}
	\caption{The expected evolution of an Option can be represented by the Weak-Value trajectory. The extremal points of the uncertainty bars (blue lines) defined at each instant $t_n$ of the evolution, represent the location of the two slits at each time. The purple lines illustrates this situation by showing which points of the graphic corresponds to the slits at a given instant of time.}
	\label{Fig4}
\end{figure}

\begin{figure}
	\centering
		\includegraphics[width=10cm, height=8cm]{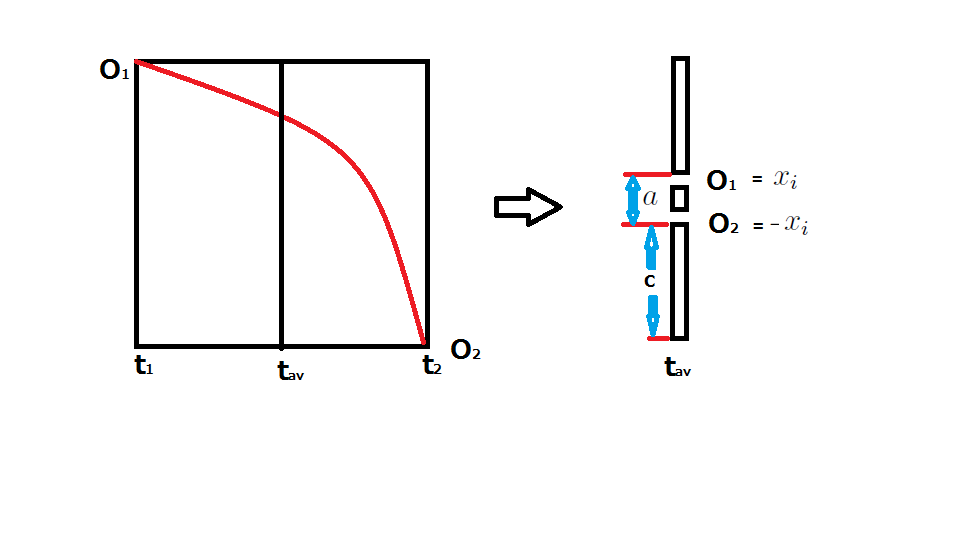}
	\caption{Application of the concept of weak-value in Quantum Finance in real situations. The figure illustrates a situation of high fluctuations for the price of an Option in some fixed interval of time. The largest price $O_1=x_i$ and the smallest price $O_2=-x_i$ appear at different times. By taking an average over the interval of time, it is possible to convert this problem to a double slit experiment. The approximation to a double slit becomes better when the interval under consideration is short. The figure also remarks that the prices are defined to be positive by considering the shift $c$. Note that the scales of the left-hand side of the figure and the right-hand side figure are different.}
	\label{Fig2}
\end{figure}

\section{THE INVERSE DOUBLE SLIT IN QUANTUM FINANCE}   \label{V}

Here we will repeat the same arguments of the previous section, but this time we will
consider two possibilities for the final state defined as $\pm x_f$ and only one possibility for the
initial state $x_i$. This can be considered as the inverse double slit problem in Quantum
Finance. Since the Kernel amplitude describes the evolution in time-reversal, then we only
have to follow the previous arguments and the only change will be in the definition given in
eq. (\ref{20}). Here we will define

\begin{equation}
\vert\phi>=\frac{1}{\sqrt{2}}(\vert x_f>+\vert -x_f>),
\end{equation}
for the final stock prices. On the other hand, we will consider $\vert\phi>=x_i$. These conditions
for the post and the pre-selected states, give us the expression

\begin{equation}  \label{200}
x_w^F(t)=\frac{<x_i\vert U(T-\tau)x^FU(\tau)\vert \psi>}{<x_i\vert U(T)\vert \psi>},
\end{equation}
instead of the result (\ref{20}). We can then make the corresponding expansion

\begin{equation}   \label{33}
x_w^F(t)=\frac{<x_i\vert U(T-\tau)x^FU(\tau)\vert x_f>+<x_i\vert U(T-\tau)x^FU(\tau)\vert -x_f>}{<x_i\vert U(T)\vert x_f>+<x_i\vert U(T)\vert -x_f>}.
\end{equation}
If we repeat the same procedures as before, then we obtain the following result
	
\begin{equation}   \label{3Amazing}
x_w^F(t)=x_i\frac{\tau}{T}+x_f\left(1-\frac{\tau}{T}\right)tanh\left(\frac{x_f}{T\sigma^2}\left(x_i+T\left(r-\frac{\sigma^2}{2}\right)\right)\right). 
\end{equation}	
Note the difference between this result and the one obtained in eq. (\ref{26}) if we exchange the
roles of $x_i$ and $x_f$. The difference is due to the fact that now we are considering the two
slits to be located at the end of the path, namely $x_f$ and not at the beginning as in the
previous case. Note the similarity of the previous result with respect to the one obtained in
eq. (\ref{14}) for the standard double slit experiment in ordinary Quantum Mechanics. However, now the weak-value trajectory is real and not complex as in the standard case of Quantum
Mechanics. In order to get the previous result, we have used the definition
	
\begin{equation}
x^\pm_w(t)=(x_i\mp x_f)\frac{\tau}{T}\pm x_f.
\end{equation}
Compare this relations with the definitions (\ref{27}). The difference is once again the fact that
the two slits corresponds to the final state. Take into account once again that the time $\tau$ is
running to the past. We can proceed now to test the result (\ref{3Amazing}). For $\tau=0$ or $t=T$, we get

\begin{equation}
x_w^F(T)=x_ftanh\left(\frac{x_f}{T\sigma^2}\left(x_i+T\left(r-\frac{\sigma^2}{2}\right)\right)\right)
\end{equation}
On the other hand, when $\tau=T$, which is equivalent to the condition $t=0$, then we get
$x_w(0)=x_i$, which is the expected result. It is also trivial to get a generic result for this case
if we introduce the odd function $f_{odd}(x_i,\pm x_f)=\pm f_{odd}(x_i, x_f)$, getting then an analogous
result to the one obtained in eq. (\ref{2Amazing}). The extension to the case of $n$-final prices defined at
a single value is also trivial. For applying this case to real situations we only have to repeat the same arguments explained in the previous section and illustrated in the Figures (\ref{Fig4}) and (\ref{Fig2}).

\section{Applications of the Weak-Value approach for the daily trading}   \label{Chopra2}

When we want to analyze the daily trading (retail trading), the previously introduced tool is remarkably useful. In the Stock market, when the amount of tradings corresponds to the usual daily average for the company under analysis, then there is not so much uncertainty in the prices and this situation is not useful for daily traders. Retail traders are trying to find volatility in the prices for getting high income and this only happens when the volume of trading is relatively high. Daily traders are then looking for stocks where the amount of people and corporations trading is unusually larger than the average. If the trading volume for some specific stock is just the average one, then the trading is being dominated by institutional traders and high frequency trading computers. Then daily traders avoid these stocks because they do not represent volatility nor any opportunity for getting some income \cite{Chopra}.
When the volume of trading is larger than the average, the prices of the Stock can either, fall or increase as can be observed from the figure (\ref{Chart}). Note the two arrows in the figure, which represent the days where the volume of trading was larger than the usual. The key point for the success of retail traders is to locate these points for any Stock and then proceed to take the appropriate decisions, namely, buying or selling. The stocks with high relative volume, trade independently of what the sector in the market is doing. These stocks are called "Alpha predators" and retail traders must focus on these stocks if they want to get high income from their investment \cite{Chopra}.    
\begin{figure}
	\centering
		\includegraphics[width=10cm, height=8cm]{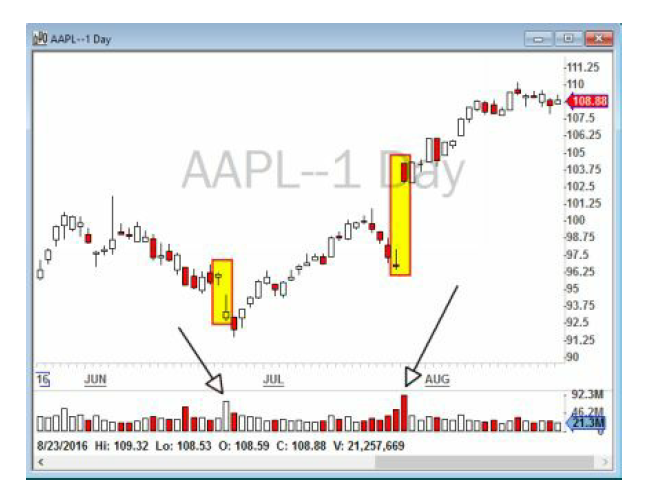}
	\caption{Daily chart for Apple corporation for summer 2016. The days with significant relative volume are marked with arrows. During those days, the price of the Stock either, dropped significantly in one case and increased significantly in the other one. The tool introduced in this paper is very useful for these kind of situations. Figure taken from \cite{Chopra}.}
	\label{Chart}
\end{figure}
The "Alpha predators" have a high-uncertainty associated to them. It is in the situations of high uncertainty that the weak-value representation of the double slit in Quantum Finance is useful in the analysis of the market. The figure (\ref{Chart}), represents a typical chart of a stock with the two arrows representing situations of high uncertainty. In the figure for example, at the times of high-uncertainty (located by the arrows), the gap between the highest prices and the lowest ones in both cases (falling price or growing price) can be represented by a double-slit and the weak-value approach proposed here is ideal. If some entrepreneur wants to develop a software able to predict realistic data, then it would be necessary to start from ideal situations as the one described in this paper and by using Machine (Deep) learning techniques, the system could learn how to make the corresponding corrections over the ideal case. We conclude that real situations in the Stock market, are described by deviations of the ideal situations. A system of artificial intelligence, wanting to learn the behavior of the market can use as a staring point the equations formulated in this paper. This is how important software development can be carried out. 

\section{Conclusions}   \label{VI}

We have found an extension of the weak-value formulation of the double slit experiment
when we consider financial Hamiltonians. In Quantum Finance, the Quantum evolution
of the system is non-unitary (no probability conservation) in general because the Hamiltonians involved are normally non-Hermitian and the $PT$ symmetry is not necessarily satisfied. Non-unitary evolutions are expected in Finance because there is never a guarantee that the probability will be preserved in time when there is a lot of information going inside and outside a system, as it is the case of a Financial market. The non-conservation of the probability does not mean that the probability itself is not well-defined. In fact, the probability is always well-defined and the total sum of probabilities at each instant of time is always one. However, the fact that there is a flow of information through the boundaries of a system (as it is the case of the Stock market), changes the normalization factor associated with the probability at each instant of time. In Quantum finance we have found that the weak-value associated to the price of an option for the double
slit experiment is normally a real number and not an imaginary quantity as in ordinary
Quantum Mechanics. In some special cases it might be possible to get complex values. A complex value would mean that besides the amplitude, there is a phase associated to the number which might be related to a complementary variable (conjugate) \cite{4}. In
this paper we analyzed two possibilities for the double slit experiment in Quantum finance.
The first one corresponds to the case where the two slits (two option prices defined at the
same time) are located at the beginning of the path. The second case corresponded to the
one where the two slits were located at the end of the path. In the analysis we have taken
into account that the evolution of the system is described by the time $\tau$ which runs backward
and not forward. This means that we have to exchange the roles of the pre-selected and
the post-selected states when we evaluate the weak-value. The Weak-value approach is useful for studying cases where there is only partial information about the prices of a Stock at a given instant of time as it is illustrated in the figure (\ref{Fig3}). If we follow real situations, we can only assign one price for an Option at a given time. However, even in these cases, the concept of double-slit in the prices of an Option, as well as the weak-value associated to it are very useful. For these cases for example, we can describe the real trajectory for the price of an Option as the Weak-value trajectory and then the uncertainty bars, defined at each instant of time, are defined as the separation between the slits as it is illustrated in Fig. (\ref{Fig4}). In addition, if we have situations with strong fluctuations as the ones illustrated by the arrows in the figure (\ref{Chart}) in connection with the "Alpha predators", then we can take an average over the intervals with high-fluctuations and convert the problem to an equivalent one where the price evolves in agreement with the Weak-value and then the uncertainty bars are connected with the distance between the highest and the lowest price (double-slit) defined through the interval over which we take the average. This case is illustrated in Fig. (\ref{Fig2}) and the chart appearing the the figure (\ref{Chart}). After average we then come back to the same case illustrated in Fig. (\ref{Fig4}). Note that the predictability of this formalism depends on the Hamiltonian under consideration. Then different financial models will provide different predictions. Since the previous formulation is general and independent of the Hamiltonian under consideration, then this formalism can
be used for testing the predictability of different models for situations where the uncertainty
in the prices is very large at a given instant (interval) of time. Then the present tools can help
us to improve the models for these kind of situations. As an example of the application of these ideas, we have worked out the ideal case represented by the Black-Scholes equation but any other equation which can be expressed as an eigenvalue problem can be used for similar analysis but different predictions. Extension to the Black-Schole equation can appear by introducing potential terms to the Hamiltonian in the form $\hat{H}=\hat{H}_{BS}+V$ and in such a case, we can enter to more realistic scenarios. Machine and Deep learning techniques for the creation of software applications can be done by starting from ideal situations as the one described in this paper. Remember that although the Black-Scholes equation has known limitations, already known by the traders, it is a good starting point for the understanding of the market. As Box famously said: "Remember that all models are wrong; the practical question is how wrong do they have to be to not be useful" \cite{BOX}.     
Other authors in the past have explored the possibility of applying Quantum Mechanical techniques for the analysis of the Black-Scholes equation in particular \cite{ref1}. However, no one of these cited authors have mentioned before the possibility of modeling the uncertainties in the prices of the Options in the market by using the double-slit approach and nobody before explained how to use the concept of weak-value for this purpose. The concept of weak-value certainly simplifies the analysis of the evolution of the prices and it will be a powerful tool for applications connected with Machine (and Deep) learning.     \\\\

{\bf Acknowledgement} I. A would like to thank to Prof. Yu Chen from The University of Tokyo for valuable discussions about this project and for providing useful bibliography for this project. I. A. is supported by the Open University of Hong Kong. C. S. is supported by IMUNAM. The authors also would like to thank the Institute of International Business and Governance of the Open University of Hong Kong, partially funded by a grant from the Research	Grants Council of the Hong Kong Special  Administrative Region, China (UGC/IDS16/17), for its support. I. A. would like to give special thanks to Miguel De Alba and Felipe Caycedo, start-ups and entrepreneurs working in Seoul-Korea, for sharing the bibliography \cite{Chopra} and for explaining how useful can be the result proposed in this paper for the development of a software.

\end{document}